\documentclass[conference]{IEEEtran}
\IEEEoverridecommandlockouts
\usepackage{cite}
\usepackage{amsmath,amssymb,amsfonts}
\usepackage{graphicx}
\usepackage{textcomp}
\usepackage{xcolor}

\usepackage{amsmath, amssymb, amsthm,algorithm}
\usepackage[latin1]{inputenc}
\usepackage{xcolor}
\usepackage{latexsym}
\usepackage{graphics,epsfig}
\usepackage{amsfonts}
\usepackage{booktabs}
\usepackage{color}
\usepackage{multirow}
\usepackage[justification=centering]{caption}
\usepackage{graphicx}
\usepackage{adjustbox}
\usepackage{kantlipsum}
\usepackage[caption=false,font=footnotesize]{subfig}
\usepackage{cases}
\usepackage{url}
\usepackage{tabu}
\usepackage{makecell}
\usepackage{soul}

\usepackage{array}
\usepackage{algpseudocode}
\usepackage[justification=centering]{caption}

\usepackage[caption=false,font=footnotesize]{subfig}
\usepackage[T1]{fontenc}
\usepackage{mwe}
\usepackage{subfig}
\usepackage{mathtools,lipsum,cuted}

\usepackage[english]{babel}
\usepackage{verbatim}

\usepackage[english]{babel}

\def\BibTeX{{\rm B\kern-.05em{\sc i\kern-.025em b}\kern-.08em
    T\kern-.1667em\lower.7ex\hbox{E}\kern-.125emX}}
\begin{document}

\title{An Experimental Study of Network Coded REST HTTP in Dynamic IoT Systems}
\author{\IEEEauthorblockN{Cao Vien Phung, Jasenka Dizdarevic and Admela Jukan}
\IEEEauthorblockA{Technische Universit\"at Braunschweig, Germany\\
Email: \{c.phung, j.dizdarevic, a.jukan\}@tu-bs.de
}}
\maketitle

\begin{abstract}
REST HTTP is the communication protocol of choice for software developers today. In IoT systems with unreliable connectivity, however, a stateless protocol like REST HTTP needs to send a request message multiple times, and it only stops the retransmissions when an acknowledgement arrives at the sender. In our previous work, we studied the usage of random linear network coding (RLNC) for REST HTTP protocol to reducing the amount of unnecessarily retransmissions. In this paper, we experimentally validate the study and analyze REST HTTP with and without RLNC in a simple testbed in dynamic IoT systems. The measurements show notable improvements in bandwidth utilization in terms of reducing the retransmissions and delay when using network-coded REST HTTP.
\end{abstract}


\section{Introduction}
Due to the large volumes of data that will be generated with Internet-of-Things (IoT) devices, -- more than $25$ billion devices reportedly by the end of $2020$ \cite{7509394}, and their technological heterogeneity, it is necessary to build communication systems and architectures that would support numerous communication techniques adoptable in different network environments. One of the most popular communication protocols in computing systems is REST HTTP (REST and HTTP stand for Representational State Transfer and Hypertext Transfer Protocol, respectively) and, despite its various critical performance shortcomings, is known as the communication protocol of choice for developers. Its popularity in the software developers community is rooted in its good performance tested in traditional cloud computing systems.

Implementations of REST HTTP in dynamic wireless sensor network environments, with uncertain network conditions are however particularly challenging. This is because a stateless protocol like REST HTTP needs to send a request message multiple times, and it only stops the retransmissions when an acknowledgement arrives at the sender. Retransmissions on the other hand are highly undesirable, due to limited resources both on the computing and communication sides of IoT. To address this issue, our previous work \cite{8756782} studied the application of REST HTTP with and without random linear network coding (RLNC), a coding technique invented in \cite{1228459}. In this paper, we studied the same protocol application in IoT systems, however, experimentally. The experiments present a few interesting findings. First, REST HTTP with network coding can reduce not only the number of unnecessary additional messages but also delay. Second, the overhead of coding and decoding time affects significantly the performance of REST HTTP with RLNC, while the overhead of traffic for responding acknowledgements for the traditional REST HTTP is shown to be higher than in case with REST HTTP with RLNC. Finally, in addition to the number of additional request messages analyzed, an experimental analysis can evaluate the protocol overhead of request messages also in bytes, which is practically relevant. While the idea of combining application layer protocols with network coding is rather new, such as adapting REST unsafe methods, used in CoAP (Constrained Application Protocol) and HTTP implementations \cite{8756782, Phungcoap}, it has not been tested yet experimentally. 

The remainder of this paper is organized as follows. Section \ref{systemdesign} presents the experimental setup scenario.  In Section \ref{exp}, we compare the performance of REST and NC\_REST  experimentally.  Section \ref{conc} concludes the paper.

\section{Experimental Setup} \label{systemdesign}
\subsection{Background and Reference Scenario} \label{background}
The concept and numerical analysis proposed in \cite{8756782, Phungcoap} are based on performing RLNC over REST and rather than sending native messages, the aim is to send coded messages. This would allow the prediction of the loss rate and subsequent modification of the number of additional messages that have to be retransmitted in dynamic scenarios with intermittent connection between a mobile IoT client device and a server edge or cloud computing node in mind. 
While the idea of reducing the number of messages sent with REST mechanism  has been put in place, compressing data in mobile edge networks has been proposed also in \cite{8624521} and \cite{AZAR2019168}. Different practices of RLNC utilization, as a well known technique for improving the communication reliability are also finding its place in novel computing paradigms for IoT based systems, as can be seen in \cite{8472148, 8790757} and its applications in optimizing data delivery and collection processes. In this paper, we instead focus on the potential network coding can have on application layer message exchange in terms of extending our REST HTTP with RLNC numerical analysis to compare the experimental one. 

Let us now briefly summarize the scenario of traditional REST HTTP (REST) and REST HTTP with network coding (NC\_REST) as earlier studied in \cite{8756782}. This summary is beneficial as we also make some small modifications from \cite{8756782} to improve the system of NC\_REST. As in previous work we start with the definition for the concept of "Seeing a packet" \cite{Jay2011} as it is necessary for this paper as well:

\textbf{Definition 1} (Seeing a packet): A node is said to have seen a packet $p_k$ if it has enough information to compute a linear combination of the form ($p_k+q$), where $q=\sum_{l>k}\delta_l p_l$, with $\delta_l\in F_q$ for all $l>k$. Thus, $q$ is a linear combination involving packets with indices larger than $k$.

In order to understand Definition $1$, we give an example: Assume that the server collects two coded packets $d_1=p_1+2p_2$  and $d_2=p_1+4p_2+5p_3+p_4$. After Gauss-Jordan elimination (GJE), we get a two new ones $d_1^, = p_1-5p_3-p_4$ and $d_2^,=p_2+\frac{5}{2}p_3+\frac{1}{2}p_4$. We observe that $d_1^,$ and $d_2^,$ follow the structure $p_k+q$ of Definition $1$. So, packets $p_1$ and $p_2$ are seen while packets $p_3$ and $p_4$ are not seen.

\begin{figure*}[!h]
  \centerline{
  \subfloat[REST HTTP without network coding]{\includegraphics[width=5.5 in, height=3.0 cm]{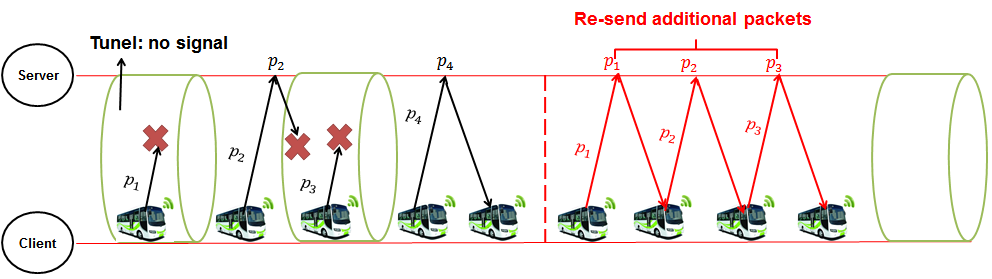}
  \label{scenarioWoNC}}
  }
  \hfil
  \centerline{
  \subfloat[REST HTTP with RLNC]{\includegraphics[width=5.5 in, height=4.1 cm, scale = 0.8]{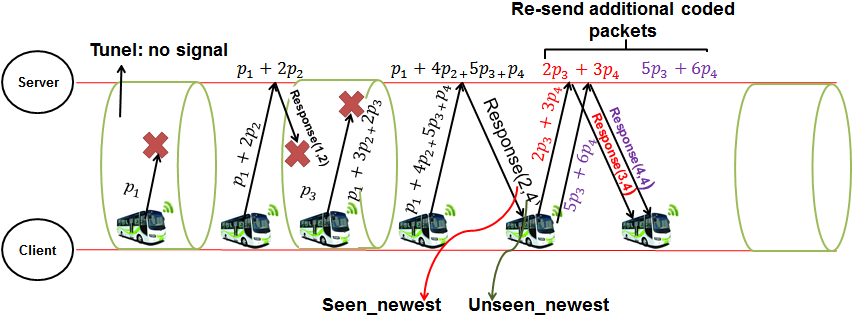}
  \label{scenarioWNC}}
  }

  \caption{Scenario of REST HTTP with and without RLNC from \cite{8756782}.}
  \label{scenario}
  \vspace{-0.3cm}
\end{figure*}

The dynamic scenario of REST HTTP with and without network coding, where connectivity issues often cause message losses, is shown in Figure \ref{scenario}, taken from \cite{8756782}. Figure \ref{scenario} shows an example where one mobile client device wants to send four request messages $p_1$, $p_2$, $p_3$ and $p_4$ related to unsafe methods (e.g. POST, PUT methods) with four different connections to one static server. To satisfy a certain reliability for this kind of system, retransmission can be performed several times after each timeout event until REST HTTP client receives its acknowledgement from the server \cite{Edstrom2012}. Nevertheless, sometimes retransmission can cause a redundancy in network traffic. As analyzed in \cite{8756782}, the traditional REST HTTP in Figure \ref{scenarioWoNC} shows that message $p_2$ is unnecessarily updated again because it has already arrived at the server, while the client cannot make sure what is happenning at the server side in unstable  connections, therefore resending it one more time and causing a bandwidth waste. As a result, to resolve this issue, a solution of REST HTTP with RLNC can be used, as presented in Figure \ref{scenarioWNC}. In this figure, we see that a new coded message includes a new native message, and this feature helps us to reduce the number of additional retransmissions. At the time of collecting the two random linear combinations $p_1+2p_2$ and $p_1+4p_2+5p_3+p_4$ (the coding coefficients are randomly chosen for the whole message), the server performs GJE on them, to find out messages $p_1$ and $p_2$ are seen and messages $p_3$ and $p_4$ are unseen (please refer to definition $1$ for concept of "Seeing a packet").  After that, the server can immediately respond to the client, even when it has not decoded the coded messages to get the desired request messages yet, i.e. response message Response($2.4$), where $seen\_newest=2$ identifies message $p_2$ as the newest seen message and $unseen\_newest=4$ identifies message $p_4$ as the newest unseen message. In order to compensate message losses, at the client side, we analyze the mentioned response message, and then redispatch two additional random linear combinations $2p_3+3p_4$ and $5p_3+6p_4$, using computation of the number of additional coded messages $R=unseen\_newest - seen\_newest=4-2=2$ (Note that after the response message Response($2,4$) arrived at the client side, messages $p_1$ and $p_2$ have been deleted from the client coding buffer, therefore they are not used in these additional linear combinations). When the client receives the response message Response($4$), that means $seen\_newest=unseen\_newest=4$ and all of original request messages have already been decoded. 

\begin{figure}[htb]
\centering
\includegraphics[width=0.5\columnwidth]{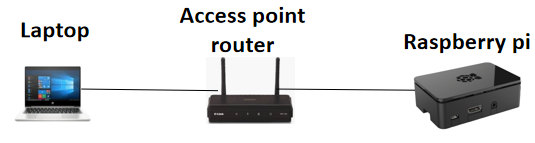}
\caption{An experimental wired network setup.}
\label{setup}
\vspace{-0.1cm}
\end{figure}

\begin{figure}[htb]
\centering
\includegraphics[width=0.5\columnwidth]{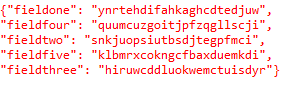}
\caption{Native POST data under a JSON file.}
\label{json}
\vspace{-0.1cm}
\end{figure}

We now present our modifications from \cite{8756782} to simplify the coding header, response message and decoding algorithm complexity in the case of high connection loss probability:
\begin{itemize}
\item For practicality, we redesign the coding header for a random linear combination: "$\overbrace{ID_{oldest},ID_{newest}}$ ; $\overbrace{Length_{ID_{oldest}},Length_{ID_{oldest}+1},...,Length_{ID_{newest}}}$ ; $\overbrace{\delta_{ID_{oldest}},\delta_{ID_{oldest}+1},...,\delta_{ID_{newest}}}$". The first part is $ID$ list (list of message identifiers) involved in a random linear combination, where $ID_{oldest}$ is the index of oldest request message and $ID_{newest}$ is the index of newest request message. The second part denotes the length list of request messages involved in a random linear combination; this field is necessary for decoding process in order for the server to prune dummy zero symbols that the client appends to the shorter request messages for the coding process. The third part shows the coding coefficient list involved in a random linear combination.
\item Response message: if $seen\_newest=unseen\_newest$, then response message only needs to contain one value of $seen\_newest$ or $unseen\_newest$. For example, response message Response($4$) in Figure \ref{scenarioWNC} means $seen\_newest=unseen\_newest=4$.
\item Buffer management method at the server side: Arrived request message will refuse the service, if the server is busy with data processing (e.g., performing GJE.), and it will be retransmitted later by the client.
\end{itemize}

As analyzed theoretically in \cite{8756782}, the number of additional request messages of REST and of NC\_REST are computed by Eq.\ref{A_WoNC}  and Eq.\ref{A_WNC}, respectively:
\begin{equation}
A_{WoNC} = \frac{N}{1-p} - N
\label{A_WoNC}
\end{equation}

\begin{equation}
A_{WNC} = \frac{N}{1-(\alpha\cdot p)} - N
\label{A_WNC}
\end{equation}
where, $N$ is the total number of request messages that the client wants to send to the server, $p$ represents the total loss probability including both the message loss on the way to the server and on the way to the client, $\alpha$ denotes the loss rate when dispatching request message to the server side.
 \begin{figure*}[ht]
  \centering
  \subfloat[Number of additional request messages]{\includegraphics[ width=4.5cm, height=4.7cm]{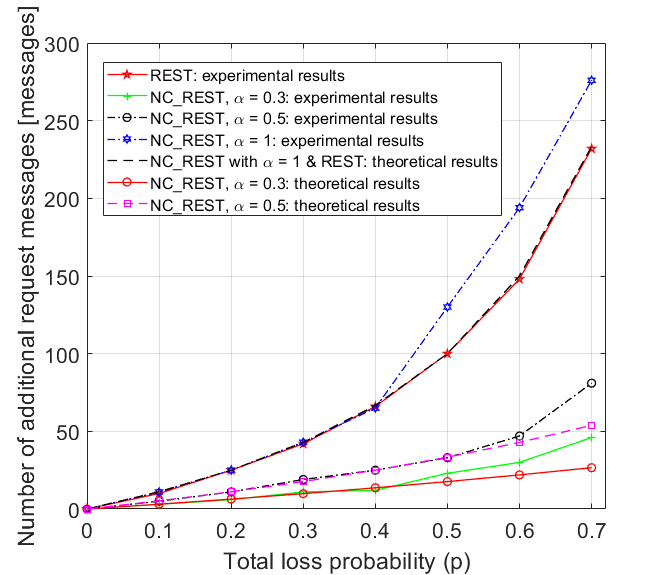}
  \label{messages_alpha03alpha05alpha1_subset5_timeout15}}
  \subfloat[Number of additional bytes of request messages]{\includegraphics[ width=4.5cm, height=4.7cm]{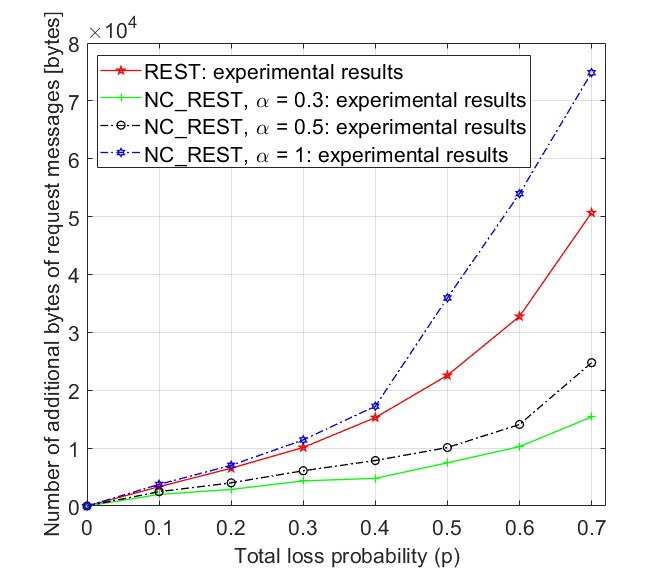}
  \label{bytes_alpha03alpha05alpha1_subset5_timeout15}}
  \subfloat[Completed time]{\includegraphics[ width=4.5cm, height=4.7cm]{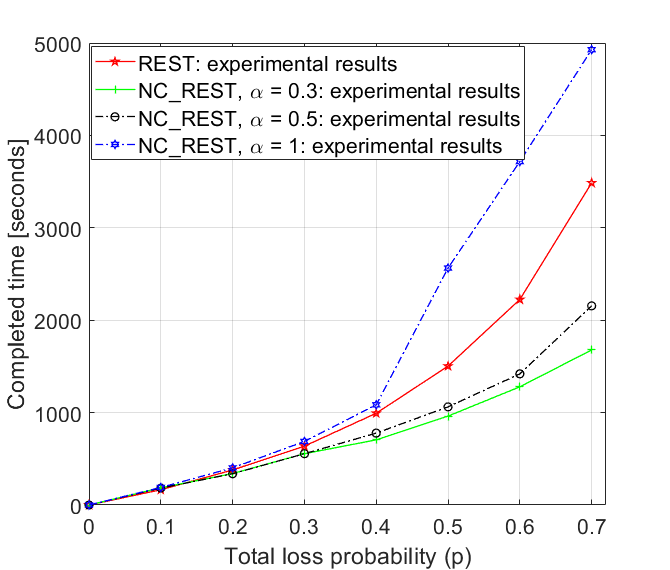}
  \label{completedtime_alpha03alpha05alpha1_subset5_timeout15}}
  \subfloat[Average decoding time]{\includegraphics[ width=4.5cm, height=4.7cm]{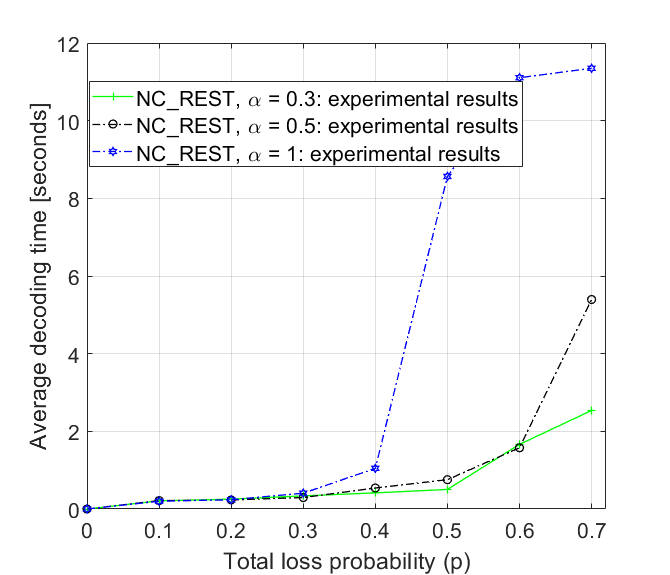}
  \label{decodingtime_alpha03alpha05alpha1_subset5_timeout15}}
  \caption{Comparison of NC\_REST and REST, where $\alpha = 0.3;0.5;1$ and subset coding buffer $SCB$ = $5$.}
  \label{alpha03alpha05alpha1subset5}
  \vspace{-0.2cm}
  \end{figure*}

 \begin{figure*}[ht]
  \centering
  \subfloat[Number of additional request messages]{\includegraphics[ width=4.5cm, height=4.7cm]{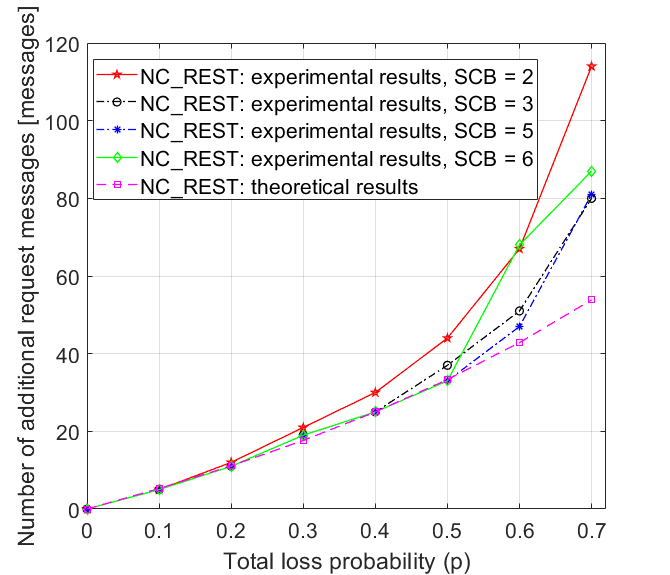}
  \label{messages_alpha05_allsubset_timeout15}}
  \subfloat[Number of additional bytes of request messages]{\includegraphics[ width=4.5cm, height=4.7cm]{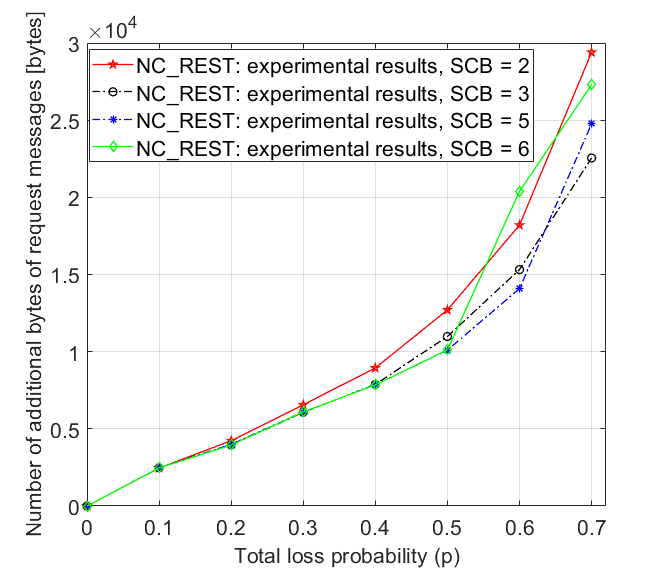}
  \label{bytes_alpha05_allsubset_timeout15}}
  \subfloat[Completed time]{\includegraphics[ width=4.5cm, height=4.7cm]{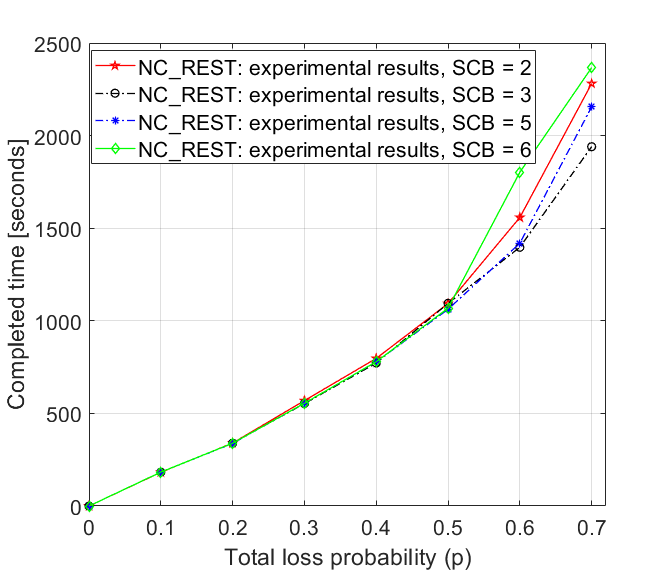}
  \label{completedtime_alpha05_allsubset_timeout15}}
  \subfloat[Avaerage decoding time]{\includegraphics[ width=4.5cm, height=4.7cm]{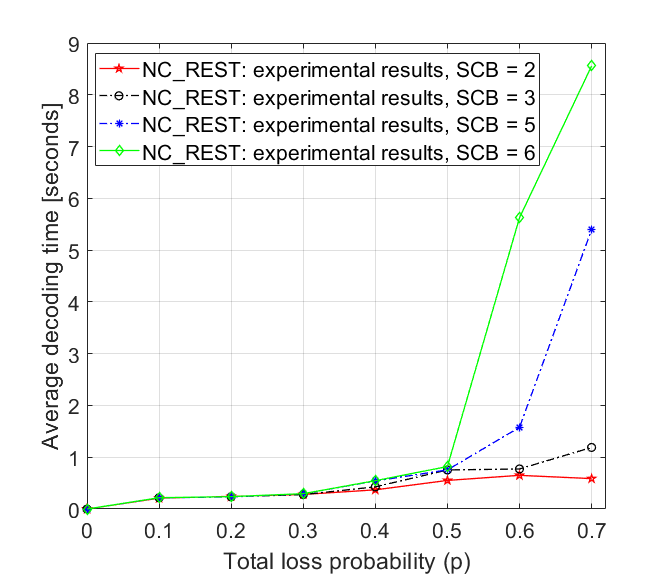}
  \label{decodingtime_alpha05_allsubset_timeout15}}
  \caption{Comparison of NC\_REST and REST, where $\alpha = 0.5$ and subset coding buffer $SCB$ = $2;3;5;6$.}
  \label{alpha05allsubset}
  \vspace{-0.2cm}
  \end{figure*}

\subsection{Experimental setup}\label{exp}
This section shows experimental wired network setup we used for client server communication with traditional REST HTTP and proposed REST HTTP with network coding. For our mobile client we used a laptop (Intel(R) Core(TM) i7-3667U CPU @ 2.00GHz) and connected it with a RaspberryPi 3 Model B+ (64-bit quad-core ARM Cortex-A53 processor @ 1.4GHz and 1 GB of RAM) as the static server running flask version 1.0.3 in python, via an access point router, as shown in Figure \ref{setup}. To simplify the setup, we decided for wired network and instead of mimicking wireless losses we emulated the link losses with Algorithm 1, until timeout expiration to get the total loss probability $p$ and loss rate $\alpha$, and then embedded them into our experimental setup. To emulate this, we show an example as follows: First, we randomly create a $matrix=[0,1,0,1,0]$ with the desired loss probability, for instance, $40\%$ of $1$, i.e. $40\%$ of connection losses until timeout expires, where $1$ denotes the connection loss until timeout expires, and $0$ denotes that there are no connection losses. As shown in Algorithm \ref{losscon}, at the client side when doing
experiments if we meet the value of $1$ from the connection loss matrix, then we will pause a time interval $t$, e.g. $t=timeout$, as if the connection was intermittent. Similarly, at the server side we pause for a long enough period of time so that the acknowledgement cannot go back to the client; when timeout expires, REST HTTP client forgets the previous one and then opens a new
connection for retransmission. Note that Algorithm \ref{losscon} should be implemented before sending the request message and the response message.

\begin{table}[]%
\caption{Average coding time in seconds with $SCB$ = $5$}
\label{Glossary}
\begin{center}
\setlength\tabcolsep{3pt}
\begin{tabular}{|c|c|c|c|c|c|c|c|}
  \hline
$p$ & 0.1 & 0.2 & 0.3& 0.4 & 0.5 & 0.6 & 0.7\\ \hline
$\alpha=0.3$ & 0.0123 & 0.0126 & 0.0130 & 0.0132 & 0.0134 & 0.0156 & 0.0179 \\ \hline
$\alpha=0.5$ & 0.0125  & 0.0125 & 0.0139 & 0.0132 & 0.0147 & 0.0145 & 0.0184\\ \hline
$\alpha=1$ & 0.0120  & 0.0121 & 0.0124 & 0.0132 & 0.0183 & 0.0200 & 0.0202\\ \hline
  
\end{tabular}
\end{center}
\end{table}

Following the REST architecture for sending POST message from the client to the server we use python request library and protocol HTTP/1.1. In order to send a native POST message in the case of REST to the Raspberry Pi, we use $message = requests.post('http://192.168.1.3:5000/api/sensors', timeout=T_o, data=Native\_POST\_Data\_with\_Coding\_Header)$, where $192.168.1.3$ is Raspberry Pi's IP address, timeout event for this experimental analysis is $T_o=15$ seconds (this value can be higher, it depends on the length of assumed connection loss) and $Native\_POST\_Data\_with\_Coding\_Header$ includes information data randomly generated under a JSON file, as shown in the example of Figure \ref{json}, and the header with the structure $\{Header: ID\}$ ($ID$ is identified for the message, e.g. message $p_1$ has $ID=1$). The length of JSON messages is chosen from small interval $[199,204]$ bytes with an arrival interval of $1$ second. 

\begin{table}[]%
\caption{Total amount of traffic in bytes for response messages to complete $100$ request messages with $SCB$ = $5$}
\label{Glossary2}
\begin{center}
\setlength\tabcolsep{3pt}
\begin{tabular}{|c|c|c|c|c|c|c|c|}
  \hline
$p$ & 0.1 & 0.2 & 0.3& 0.4 & 0.5 & 0.6 & 0.7\\ \hline
$REST, \alpha=0.3$ & 1702 & 1861 & 2052 & 2339 & 2672 & 3262 & 4176 \\ \hline
$NC\_REST,\alpha=0.3$ & 1599  & 1606 & 1626 & 1629 & 1737 & 1767 & 2042\\ \hline
$REST,\alpha=0.5$ & 1670  &  1796 & 1925 & 2131 & 2385 & 2734 & 3422 \\ \hline
$NC\_REST,\alpha=0.5$ &  1605 & 1622  & 1681 & 1688 & 1713 & 1777 & 2011 \\ \hline
\end{tabular}
\end{center}

\end{table}

For requests in case of NC\_REST implementation, before sending data, we perform coding to generate coded data using the source code of project Nayuki \cite{ProjectNayuki} over the finite field $\mathbb{F}_{2^8}$, where coding coefficients are randomly chosen for the whole message, as it was the case in our numerical analysis, and each coded message is only allowed to generate random linear combiantions with the number of native request messages limited by the subset coding buffer value ($SCB$) (we design $SCB$ so to limit the number of native request messages permitted to code in a random linear combiantion); this parameter is important because we cannot combine all request messages in the coding buffer, since it would lead to a high delay. The coded data is added the coding header following the structure defined in section \ref{systemdesign}.  After that, we  send the coded data similarly as in the case of traditional REST. A request message is only deleted from our client coding buffer implementation, if its $ID$ is less than or equal to the newest seen $ID$ $seen\_newest$ in the acknowledgement. We only send the number of additional request messages, if the client receives the response message that contains $unseen\_newest-seen\_newest>0$ (Note that in our experimental implementation we use stop-and-wait mechanism, so the parameter $r\_ID$ in \cite{8756782} is unnecessary to consider.)

For GJE for decoding over the finite field $\mathbb{F}_{2^8}$ we also used Nayuki \cite{ProjectNayuki} code. At server side, we design a decoding matrix to store the coding coefficients and a decoding buffer to store native and coded data. To know if a coded message is linearly dependent or independent, we only need to perform GJE on the decoding matrix. If linearly dependent, the coded one will be removed from the decoding buffer. Else, GJE will be performed on the whole arrived coded data. The response message includes two types of structure: $\{"Number": "seen\_newest"\}$ if $seen\_newest=unseen\_newest$ and $\{"Number": "seen\_newest,unseen\_newest"\}$ if $seen\_newest\neq  unseen\_newest$, where $seen\_newest$ and $unseen\_newest$ are found out after peforming GJE on the decoding matrix. Note that, we can immediately send the response message, even when desired request messages have not been decoded yet. The dummy zero symbols will be removed by using the coding header sent from the client after each original request message is decoded. A message is deleted from the decoding buffer only when it is already decoded and its $ID$ is less than $ID_{oldest}$ sent from the client (The coding coefficients in the decoding matrix related to that message are also deleted.).

Let us consider $100$ request messages are sent to the server. The number of retransmissions is not limited. In the dynamic environments with unreliable connections, we use REST HTTP with stop-and-wait mechanism because sending multiple requests at the same time would worsen the bandwidth utilization. The client buffer is able to store the number of unsuccessfuly sent request messages. The server only refuse an arrived message, if it is busy; this means that the message will not be stored in the server buffer and be responded to the client, instead it will be retransmitted later by the client after timeout expires.

\begin{algorithm}\caption{Emulation of the connection loss until timeout expires}
\begin{algorithmic}[1]
\Procedure{Solve}{$Connection\_loss\_matrix$} 
\If{we meet the value of $1$ from the connection loss matrix $Connection\_loss\_matrix$}
\State Pause ($t$), e.g. $t=timeout$
\EndIf
\EndProcedure
\end{algorithmic}
\label{losscon}
\end{algorithm}

\section{Measurements and Results}\label{exp}

In this section, we compare the performance of REST and NC\_REST experimentally. For theoretical results, Eq.\ref{A_WoNC} and Eq.\ref{A_WNC}  are used for the number of  additional requests of REST  and NC\_REST, respectively. Figure \ref{alpha03alpha05alpha1subset5} shows the theoretical and experimental results of NC\_REST and REST. The number of additional requests is shown in Figure \ref{messages_alpha03alpha05alpha1_subset5_timeout15}, while Figure \ref{bytes_alpha03alpha05alpha1_subset5_timeout15} shows the number of additional bytes of request messages. Completed time in seconds for $100$ request messages is shown in Figure \ref{completedtime_alpha03alpha05alpha1_subset5_timeout15} and average decoding time in seconds at the server side is shown in Figure \ref{decodingtime_alpha03alpha05alpha1_subset5_timeout15}, where the loss rate $\alpha=0.3;0.5;1$ and $SCB=5$. For REST, to simplify graphs, we only show one case of REST on average with values  $\alpha = 0.3$, $\alpha = 0.5$ and $\alpha = 1$. The theoretical results of REST and NC\_REST with $\alpha = 1$ in Figure \ref{messages_alpha03alpha05alpha1_subset5_timeout15} are the same because all messages lost are from the client side, so we do not have any benefit from using network coding in this case.

In Figure \ref{messages_alpha03alpha05alpha1_subset5_timeout15}, for total loss probability $p$ in interval $[0.1;0.4]$, the experimental results of NC\_REST with all values $\alpha$ match the theory. However, when $p \geq 0.5$, the number of additional request messages of NC\_REST obtained by the experiments is larger than what can be obtained from the theoretical results. For example, the experimental number of additional request messages equals $46$ messages for $p=0.7$ and $\alpha = 0.3$ while the theoretical result is about $26$ messages. For $p=0.7$ and $\alpha = 0.5$, the experimental number of additional request messages is $81$ messages while the theoretical result is about $53$ messages. For $p=0.7$ and $\alpha = 1$, the experimental number of additional request messages equals $276$ messages  while the theoretical result is about $233$ messages. This difference can be due to several reasons. First, the theoretical analysis did not take into account retransmission of request messages due to server refusing the service when busy. This happens when the total loss probability $p$ is high, i.e., $p \geq 0.5$, therefore causing high decoding complexity and time leading to retransmitting request message mulitple times. Second, the theoretical analysis cannot take into account some request messages retransmitted due to linearly dependent coded messages. Third, theory does not consider that $SCB$ and the number of request messages $N$ are limited, so many new coded messages do not include new native messages, leading to some request messages unnecessarily transmitted or retransmitted when the total loss probability is high. In Figure \ref{messages_alpha03alpha05alpha1_subset5_timeout15}, we can see that the number of additional request messages for REST obtained by the experiments corresponds to theory. For instance, the experimental and theoretical number of additional request messages equal $148$ and $150$, respectively, for $p=0.6$. The reason is because REST is simple and there are no unwanted factors that impact on its performance as the case of NC\_REST. 

The experimental results also validate the theoretical analysis that the lower the loss rate $\alpha$ is, the higher the benefits of NC\_REST are. In Figure \ref{bytes_alpha03alpha05alpha1_subset5_timeout15} with $p=0.5$, compared with NC\_REST with $\alpha = 0.3$, we can see the experimental number of additional bytes of request messages of NC\_REST with $\alpha = 0.5$ increases $35.420 \%$ and with $\alpha = 1$ increases $382.062 \%$. For our theoretical and experimental analysis, compared with REST, NC\_REST often reduces the number of additional request messages. For experimental example of Figure \ref{bytes_alpha03alpha05alpha1_subset5_timeout15}, we only consider a small total loss probability $p=0.2$, REST has to resend $6526$ bytes, but NC\_REST only resends $2862$ bytes for $\alpha = 0.3$ and $3999$ bytes for $\alpha = 0.5$. For the experimental case of $\alpha = 1$, NC\_REST has worse results than REST in term of the number of additional request messages (see Figure \ref{messages_alpha03alpha05alpha1_subset5_timeout15}), number of additional bytes of request messages (see Figure \ref{bytes_alpha03alpha05alpha1_subset5_timeout15}) and completed time (see Figure \ref{completedtime_alpha03alpha05alpha1_subset5_timeout15}) because $\alpha=1$ means all lost messages are from the client side while NC\_REST always only resends for the lost request messages. Although some cases of NC\_REST with $\alpha = 1$ and REST have the same experimental number of additional request messages, the experimental number of additional bytes of request messages of NC\_REST  is more than of REST because NC\_REST consumes the amount of traffic for the coding header. For example of $p=0.2$ in Figure \ref{bytes_alpha03alpha05alpha1_subset5_timeout15}, NC\_REST with $\alpha=1$ and REST additionally transmit $25$ request messages equivalent to $7034$ additional bytes and only $6526$ additional bytes, respectively.

In Figure \ref{completedtime_alpha03alpha05alpha1_subset5_timeout15}, we see that NC\_REST can decrease reconnection time by reducing the number of additional bytes of request messages, except for NC\_REST with $\alpha = 1$. For instance, at the point of $p=0.5$, the completed time for $100$ request messages is $962$ seconds for NC\_REST with $\alpha = 0.3$, $1062$ seconds for NC\_REST with $\alpha = 0.5$, $1503$ seconds for REST and $2564$ seconds for NC\_REST with $\alpha = 1$.  Figure \ref{decodingtime_alpha03alpha05alpha1_subset5_timeout15} shows the average decoding time, one of the factors affecting the performance of NC\_REST. Observe that NC\_REST with $\alpha = 0.3$ has the lowest average decoding time and NC\_REST with $\alpha = 1$ has the highest average decoding time. That is the reason why NC\_REST with $\alpha = 0.3$ achieves the best performance in term of the number of additional request messages, number of additional bytes of request messages and completed time for $100$ request messages. The high decoding time of NC\_REST with $\alpha = 1$ is also one of the main factors causing the significant difference about the number of additional request messages, number of additional bytes of request messages and completed time compared with REST because it can cause many arrived request messages to be refused by the server and retransmitted later by the client and delay the completed time. We observe that the average coding time of loss rate $\alpha=0.3$, $\alpha=0.5$, and $\alpha=1$ with $SCB=5$, as shown in Table \ref{Glossary}, is quite small, therefore it does not have a huge impact on the performance of NC\_REST. With the theoretical and experimental results, we can conclude that in most cases NC\_REST outperforms REST, except the case of NC\_REST with $\alpha = 1$.

Table \ref{Glossary2} shows the total amount of traffic in bytes for response messages to complete $100$ request messages with $SCB$ = $5$. We can observe that NC\_REST reduces not only the extra traffic of request messages but also the traffic overhead of acknowledgements. This is because REST\_NC reduces the number of messages exchanged between the client and server.

Figure \ref{alpha05allsubset} shows the theoretical and experimental results, where the loss rate $\alpha=0.5$; and $SCB=2,3,5,6$. We observe that in Figure \ref{messages_alpha05_allsubset_timeout15} the experimental number of additional request messages of NC\_REST with $SCB=2$ is the highest for most cases of $p$, except $p=0.1$, because as aforementioned, the small subset coding buffer value $SCB$ is the reason that many newly coded messages cannot be included in new native messages, therefore leading to many linearly dependent coded messages unnecessary when the total loss probability $p$ is high. In the total loss probability interval $p=$ $[0.1;0.4]$ of Figure \ref{messages_alpha05_allsubset_timeout15}, the experimental results of number of additional request messages of NC\_REST with $SCB=3;5;6$ are the same because the total loss probability $p$ is not so high, so a small subset coding buffer value $SCB=3$ is enough to avoid the problem of $SCB$ limited as discussed above. However, for the high total loss probabilities $p \geq 0.5$, $SCB=5$ may be a good choice for reducing the number of additional request messages, while $SCB=6$ is not. The reason is that although $SCB=6$ can avoid the problem of $SCB$ limited, its decoding time is quite high (this is the drawback of NC\_REST sometime affecting the quality of service of IoT systems). For example of $p=0.7$ in Figure \ref{decodingtime_alpha05_allsubset_timeout15}, the average decoding time of NC\_REST with $SCB=5$ is $5.394$ seconds while that one of  NC\_REST with $SCB=6$ is up to $8.556$ seconds.

To clearly understand the benefit of NC\_REST in selecting $SCB$, we analyze Figure \ref{bytes_alpha05_allsubset_timeout15}. For a small total loss probability $p=0.1$, NC\_REST with all subset coding buffer values  has no significant difference, we can choose the lowest $SCB=2$ with $2458$ additional bytes of request messages and $182$ seconds to complete $100$ request messages because decoding time is quite low (see Figure \ref{decodingtime_alpha05_allsubset_timeout15}), we do not consume much traffic for coding coefficients. Similarly, for the total loss probability interval $p=[0.2;0.4]$ of Figure \ref{bytes_alpha05_allsubset_timeout15}, $SCB=3$ should be selected. For $p=0.5$ and $p=0.6$, in Figure \ref{bytes_alpha05_allsubset_timeout15} the best choice is $SCB=5$. Nevertheless, for a high total loss probability $p=0.7$, we should choose $SCB=3$ because although the number of additional request  messages of $SCB=3$ and $SCB=5$ is the same, the average decoding time of $SCB=5$ is quite high, about $5.394$ seconds, compared with that of $SCB=3$, about $1.189$. In general, in term of completion time in Figure  \ref{completedtime_alpha05_allsubset_timeout15}, $SCB=3$ will be the best choice in most cases. Note that at the total loss probability of $p=0.6$ in Figure \ref{messages_alpha05_allsubset_timeout15}, the experimental number of additional request messages of  $SCB=2$ and $SCB=6$ is similar, but the corresponding experimental number of additional bytes of request messages of $SCB=2$ in Figure \ref{bytes_alpha05_allsubset_timeout15} is lower because the case of $SCB=6$ takes much more network traffic for coding header. In summary, $SCB$ is an important parameter affecting to the performance of NC\_REST. On the other hand, Figure \ref{messages_alpha05_allsubset_timeout15} leads to an interesting conclusion that NC\_REST with experimental results of $SCB=2$ and NC\_REST with theoretical results are the worst and best case, respectively, of the number of additional request messages of NC\_REST. 

\section{Conclusion}\label{conc}
In this paper, we analyzed experimentally the traditional REST HTTP and REST HTTP with network coding in terms of traffic sent and communication time for the dynamic IoT systems. Our experiential analysis can be used to validate the theoretical results and show that REST HTTP with network coding can improve not only bandwidth utilization but also communication delay. In future work, we both need to extend the theoretical model (to include analysis of coding buffer and subset coding buffer) as well as extend the experimental setup with the limitations of coding buffer to understand better the performance of REST HTTP with network coding.

\section*{Acknowledgment}
This work has been partially supported by the mF2C project funded by the European Union's H2020 Research and Innovation programme, under grant agreement 730929.

\bibliographystyle{IEEEtran}
\bibliography{nc-rest}

\end{document}